\PassOptionsToPackage{svgnames,dvipsnames}{xcolor}

\documentclass{vgtc}                          

\graphicspath{{figures/}{pictures/}{images/}{./}} 

\usepackage{times}                     

\newcommand{\dash}{\textsc{Dash}}

\usepackage{mathptmx}                  
\usepackage{tabularx}
\usepackage{tabu}                      
\usepackage{booktabs}

\newcommand{\pheading}[1]{\noindent\textbf{#1}}

\onlineid{1068}

\vgtccategory{Research}

\vgtcinsertpkg

\title{\dash: A Bimodal Data Exploration Tool for \\Interactive Text and Visualizations}

\author{Dennis Bromley\thanks{dbromley@tableau.com} %
\and Vidya Setlur\thanks{vsetlur@tableau.com}}
\affiliation{\scriptsize Tableau Research}

\teaser{
  \centering
  \includegraphics[width=\linewidth]{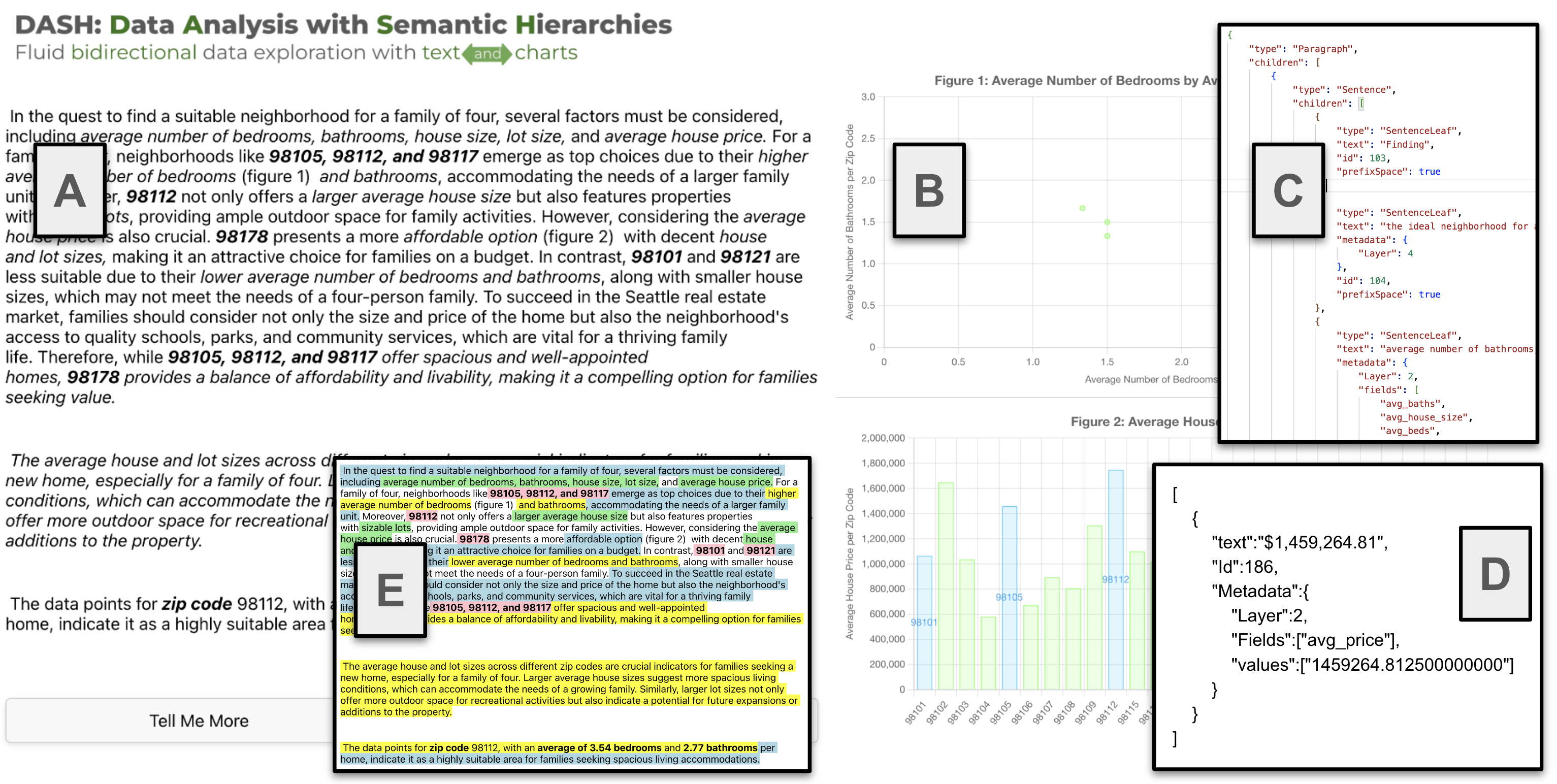}
  \caption{
    \dash{} tool's interface for supporting textual and visual data analysis. (A, B) Text and charts encoded with \dash~semantic metadata. (C) A JSON representation of semantic metadata, including the semantic level, the data field, and the data value. (D) The \dash~data-exchange JSON packet comprises the interactive text, its metadata, and identifiers that link the textual narrative to specific data points. (E) \dash~semantic level assignment using the Lundgard \textit{et al.} \cite{lundgard2021accessible} color encoding.
  }
  \label{fig:teaser}
}

\abstract{
  Integrating textual content, such as titles, annotations, and captions, with visualizations facilitates comprehension and takeaways during data exploration. Yet current tools often lack mechanisms for integrating meaningful long-form prose with visual data. This paper introduces \dash, a bimodal data exploration tool that supports integrating semantic levels into the interactive process of visualization and text-based analysis. \dash~operationalizes a modified version of Lundgard et al.'s semantic hierarchy model that categorizes data descriptions into four levels ranging from basic encodings to high-level insights.
By leveraging this structured semantic level framework and a large language model's text generation capabilities, \dash~enables the creation of data-driven narratives via drag-and-drop user interaction.
Through a preliminary user evaluation, we discuss the utility of \dash's text and chart integration capabilities when participants perform data exploration with the tool. 

} 

\keywords{Semantic levels, LLMs, text generation.}

\usepackage[svgnames,dvipsnames]{xcolor}

\begin{document}


\firstsection{Introduction}

\maketitle


The interplay between text and visualizations is an important aspect of enhancing comprehension during data exploration~\cite{segel2010narrative,kosara2013storytelling}. Research underscores the critical role of text in the reader's interpretation of visualizations, where it serves to summarize statistical attributes, explain insights, and offer contextual insights~\cite{kim2021towards}. Effective textual descriptions not only reinforce the visual elements of a chart but also improve reader engagement and trust~\cite{stokes2022striking,kong2019trust}. However, existing tools often fall short of providing robust support for authoring text alongside visualizations, typically offering limited solutions for titles and chart/text alignment~\cite{liu2023autotitle,kim2023emphasischecker}. To explore new ways of supporting interactive data exploration using \textit{both} text and visualizations, we introduce \dash~(Data Analysis using Semantic Hierarchies), a tool that leverages the integration of semantic levels in text for data analysis~\cite{lundgard2021accessible}. The tool provides large language model (LLM)-generated textual content through direct interactions with the visualization and vice-versa, supporting a \textit{bimodal} composition process that is reflective of the underlying data semantics. In Figure~\ref{fig:teaser}, (A) shows LLM-generated interactive text that contextualizes data points with semantic metadata, while (B) displays the corresponding visual data through charts. (C) outlines the JSON metadata representation that describes the semantic levels and data fields, (D) illustrates a JSON packet that facilitates bimodal/bidirectional interactive data exploration and manipulation of the narrative in real-time, and (E) shows the semantic level assignment using the Lundgard and Satyanarayan~\cite{lundgard2021accessible} color encoding. A preliminary user evaluation of \dash~ indicates utility in integrating automated and manual authoring capabilities to enhance the overall data exploration process.

\section{Related Work}
Our work is informed by research areas that explore the interplay of text and visualizations, organized into the following themes:

\subsection{Typologies and taxonomies for analytical tasks}
In the research space of visual analytics and data visualization, efforts have been made to understand and map users' data-exploration interactions to their underlying motivations and analytical goals. Brehmer and Munzner's typology of visualization tasks provides a foundational framework for this approach, offering a way to classify and interpret user actions within data visualizations~\cite{brehmer2013multi}. 

Complementing this work, Amar and Stasko developed a taxonomy categorizing the different types of questions users ask when engaging with visualizations, providing a user-centered perspective on the design and evaluation of visualization tools~\cite{amar2004}. Shneiderman offers a practical sequence for interactive visualization design, emphasizing the user's path from an overview to specifics~\cite{Shneiderman1996}. Thomas and Cook expand upon these ideas, proposing a broad framework for the visual analytics process that aligns with analytical reasoning and supports the development of interactive systems~\cite{Thomas2005IlluminatingTP}. Further refining these taxonomies, Tory and Moller address the task-based design of visualization tools, promoting an understanding of users' tasks as a central design principle~\cite{tory2004}. Gotz and Zhou examine how visual analytics tools support users in forming insights and making decisions, contributing to the understanding of how people derive value from visual data exploration~\cite{gotz2009}.

Recent work has sought to apply these theoretical constructs more directly to user interaction data. Lundgard and Satyanarayan's taxonomy~\cite{lundgard2021accessible} further refine this approach by breaking down the semantic content conveyed by visualizations into a hierarchy of comprehensible levels. This model starts at the elemental and encoded properties of the data and extends to higher-order cognitive and contextual insights, providing a structured methodology to assess and design visualization and textual description interplay. Taking inspiration from the aforementioned taxonomies, \dash~operationalizes the semantic hierarchy proposed by Lundgard and Satyanarayan, encoding textual and visual content with metadata that represents different levels of data analysis, from basic data representation to integrated domain knowledge. Users can then engage in interactive data exploration through both the generated text and visualizations.

\subsection{The role of text in visual analytics}
There is a growing consensus within the research community that text should be treated as co-equal to visualization and can significantly enhance data comprehension. 
Kim et al.~\cite{kim2021towards} conducted a study on how readers interpret line charts with accompanying captions, finding that coherence between textual and visual elements enables readers to glean insights more effectively. This coherence is particularly useful when both text and visuals emphasize the same key features, enhancing the overall understanding of the data presented. Research by Stokes et al.~\cite{stokes2022striking} highlights that users show a preference for visualizations that are heavily annotated as opposed to those with sparse or no text, suggesting that detailed annotations can reduce cognitive load and clarify complex datasets. This premise is further elaborated by Quadri et al.~\cite{quadri2024you} and Fan et al.~\cite{fan2024understanding}, who investigate how textual details affect comprehension and the perception of thematic maps, revealing that text can significantly influence how users interpret spatial and statistical data.

Ottley et al.~\cite{ottley2019curious} explore how text annotations can affect perceptions of bias and credibility of the presented data. Their findings emphasize that text not only serves to explain and summarize data but also shapes trust and reliability attributed to the visualizations by the viewers. Stokes et al.~\cite{stokes2023role} further contribute to this topic by examining how different types of semantic content impact reader takeaways, underscoring the nuanced role that text plays in data visualization environments. Our work continues this line of research by exploring how interactive and dynamically generated text can complement the visual analysis process. \dash~allows for the real-time adaptation of both text and visualizations as users interactively navigate across various semantic levels of text description.

\subsection{Visualization and text tools}
Research focusing on tools that support the integration of visualization and text has explored techniques for the creation, interpretation, and enhancement of visualizations with textual elements. He et al.~\cite{he2024leveraging} surveyed the utilization of LLMs for crafting narrative visualizations, underscoring the potential of AI in bolstering the narrative aspects of data visualization. Other tools, such as AutoTitle~\cite{liu2023autotitle}, an interactive title generator, and Vistext~\cite{tang2023vistext}, a benchmarking tool for semantically rich chart captioning, demonstrate novel capabilities in automatic text generation.

VizFlow~\cite{sultanum2021} has shown the effectiveness of facilitating interaction between authors and readers by dynamically connecting text segments to corresponding visual elements, enriching the storytelling experience. Similarly, systems such as Kori~\cite{latif2021kori} provide an interactive environment for synthesizing text and charts within data documents, emphasizing seamless integration for enhanced communication. Tools such as EmphasisChecker~\cite{kim2023emphasischecker}, DataDive~\cite{kim2024datadive}, and FigurA11y~\cite{singhfigura11y} focus on specific aspects such as guiding chart and caption creation, aiding readers in contextualizing statistical statements and supporting the crafting of accessible scientific text, respectively. Recent advances have also been made with systems like SciCarpenter~\cite{hsu2024scicapenter}, which supports the composition of scientific figure captions using AI-generated content, reflecting a broader trend toward integrating generative AI into the visualization process. Basole and Major~\cite{basole2024generative} discuss how generative AI tools offer creative assistance and automation throughout the visualization workflow, marking a shift towards a ``human-led, AI-assisted" paradigm. This premise includes everything from user and design requirements to data preparation and insight augmentation.

\dash~contributes to this research by adopting a mixed-initiative approach that leverages an LLM to enhance interactivity and semantic coherence between text and visualizations. The tool employs a semantic framework, allowing for interaction and bidirectional manipulation of both text and visual elements in real time. 

\section{\dash~Tool}
\subsection{Motivation}
Text and visual modalities each excel at different aspects of data analysis; visual charts compress large amounts of data into data-rich images, while news agencies, blogs, and even this article are \textit{text-first} documents that emphasize higher level data-analysis concepts such as inter-data relationships, conceptual discussions, and speculative narratives. This chart/text split between low-level and high-level semantic analysis inspired the design of the \dash~tool.

\subsection{Applying semantic levels for data exploration}
Lundgard and Satyanarayan~\cite{lundgard2021accessible} devised a four-level data analysis model where analysis levels increase in knowledge integration and semantic granularity from basic chart information such as axis-ranges and encodings to domain-specific insights and broader multi-domain narratives. These four semantic levels provided additional inspiration around \textit{when} to use \textit{which} modality and how to craft a compelling bimodal data exploration narrative. Therefore, we propose a similar four-level (Semantic Levels 1-4, hereafter referred to as $L1$-$L4$) data-analysis semantic hierarchy specifically designed to be operationalizable in real-world tools (Table \ref{tab:semantic_levels}). The \dash~tool utilizes this semantic hierarchy to help create a fluid data analysis experience where text and charts are both first-class concepts, each leveraging their own strengths.

\noindent
\begin{table}[htb]
    \centering
    \begin{tabularx}{\columnwidth}{@{}c>{\raggedright\arraybackslash}Xc>{\raggedright\arraybackslash}X@{}}
        \toprule
        \textbf{Level} & \textbf{Role} & \textbf{Modality} & \textbf{Example} \\ 
        \midrule
        4 & Insights and Integration of Domain Knowledge & Text & ``The steady increase in CEO pay may be playing a role in the recent uptick in investor activism.'' \\
        3 & Relationships among data \& statistics & Text, Chart & Correlations, clustering, outliers, and trends. \\
        2 & Statistics & Text, Chart & STDEV(profit) \\
        1 & Base Data & \textasciitilde Text, Chart & Rows and columns of a SQL database \\
        \bottomrule 
    \end{tabularx}
    \caption{Semantic levels for data analysis, with preferred modalities and examples at each level. Higher levels refer to higher-level semantic abstraction and knowledge integration. The \textit{\textasciitilde Text} on Level 1 modality indicates that text is often used to present individual data values but is typically not used for data presentation.}
    \label{tab:semantic_levels}
\end{table}

\subsection{Implementation}
The \dash~tool is implemented using Typescript~\cite{Typescript} and React~\cite{React} and consists of four primary pieces: a \dash~metadata definition comprising information on the data field, specific data values, semantic level, and the user-facing text (Figure~\ref{fig:teaser}D), a generative LLM (OpenAI API model \textit{gpt-4-turbo-preview)}, Slate.JS~\cite{SlateJS}, a text presentation library that maintains text-associated \dash~metadata and renders the text appropriately, 
and the chart library Chart.JS~\cite{Chartjs}.

The \dash~tool is initialized with 1) a dataset (our prototype used Seattle real estate data~\cite{SeattleRealestateData})
2) a natural language (NL) description of the dataset, 3) an NL description of the analytical end goal (in our case, to find a home for a family of four), and 4) a description of the \dash~metadata format and instructions on how to assign the different semantic levels to the LLM's response text. The LLM produces the final narrative text formatted as a tree-like hierarchy where each node in the tree contains metadata concerning semantic level (`Layer' in the Figure~\ref{fig:teaser}D JSON code) and associated data fields and values (Figure \ref{fig:teaser}C). This tree structure maintains the text organization (\textit{Paragraph} $\to$ \textit{Sentence} $\to$ \textit{SentenceLeaf}) and is necessary for consumption by the text rendering component. 

The text rendering component provides rendering callbacks to format the text based on the text's metadata. LLM response is shown in Figure~\ref{fig:teaser}E formatted with color-coded semantic levels ($L1$ (\textcolor{magenta}{pink}), $L2$ (\textcolor{ForestGreen}{green}), $L3$ (\textcolor{Goldenrod}{yellow}), and $L4$ (\textcolor{blue}{blue}).) When text is dragged from either the 
text or chart components of the \dash~interface, we recover the text's underlying metadata from the component. While text is assigned a semantic level by the LLM, chart data in this prototype is assigned to L2 because chart data is semantically higher than L1 but lower than L4. More expressive charts with analytical overlays, such as clustering and correlation, could comprise L3 data. From these data, whether sourced from text or chart, we construct the 
drag-and-drop JSON object shown in Figure \ref{fig:teaser}D and store the object in browser memory. The object can then be dropped anywhere within \dash, independent of where it came from, a crucial aspect of \dash's bidirectional data flow. 

\textit{Tell Me More}, \textit{Show Me More}, and charts (Figure \ref{fig:dash-interactivity}) are all capable of 
receiving the JSON object above via mouse drag-and-drop. \textit{Show Me More} creates a new chart with the metadata from the dropped JSON object; this metadata details the data fields and values to be charted. Inspired by tools like Polaris~\cite{stolte2002polaris} and ShowMe~\cite{mackinlay2007show}, one field produces a bar chart; two fields produce a scatter plot, specific data values produce reference lines, and specific zip codes highlight the indicated mark. If the JSON object is dropped directly onto an existing chart, the chart is updated in the same manner, possibly upgrading a bar chart to a scatter plot if a new field is added. If the drag-and-drop JSON object is dropped on \textit{Tell Me More}, \dash~re-queries the LLM for further discussion about the fields and values in the object. To produce a narrative style, \dash~produces semantically complementary responses by offering high-level ($L3$ \& $L4$) analytical responses to low-level ($L1$ \& $L2$) data observations and low-level data-centric responses to high-level observations. As shown in Figure \ref{fig:dash-interactivity}, this semantic-level-aware, source-agnostic data flow facilitates data exploration between \dash's components, including text-to-text, text-to-chart, chart-to-text, and chart-to-chart interactivity. While the \dash~interface employs drag-and-drop, the framework supports any affordance that allows the user to interactively direct data from one data-rendering modality to another.


\subsection{Interface and Interactivity Behavior}
\begin{figure*}[ht]
\centering
\includegraphics[width=0.83\linewidth]{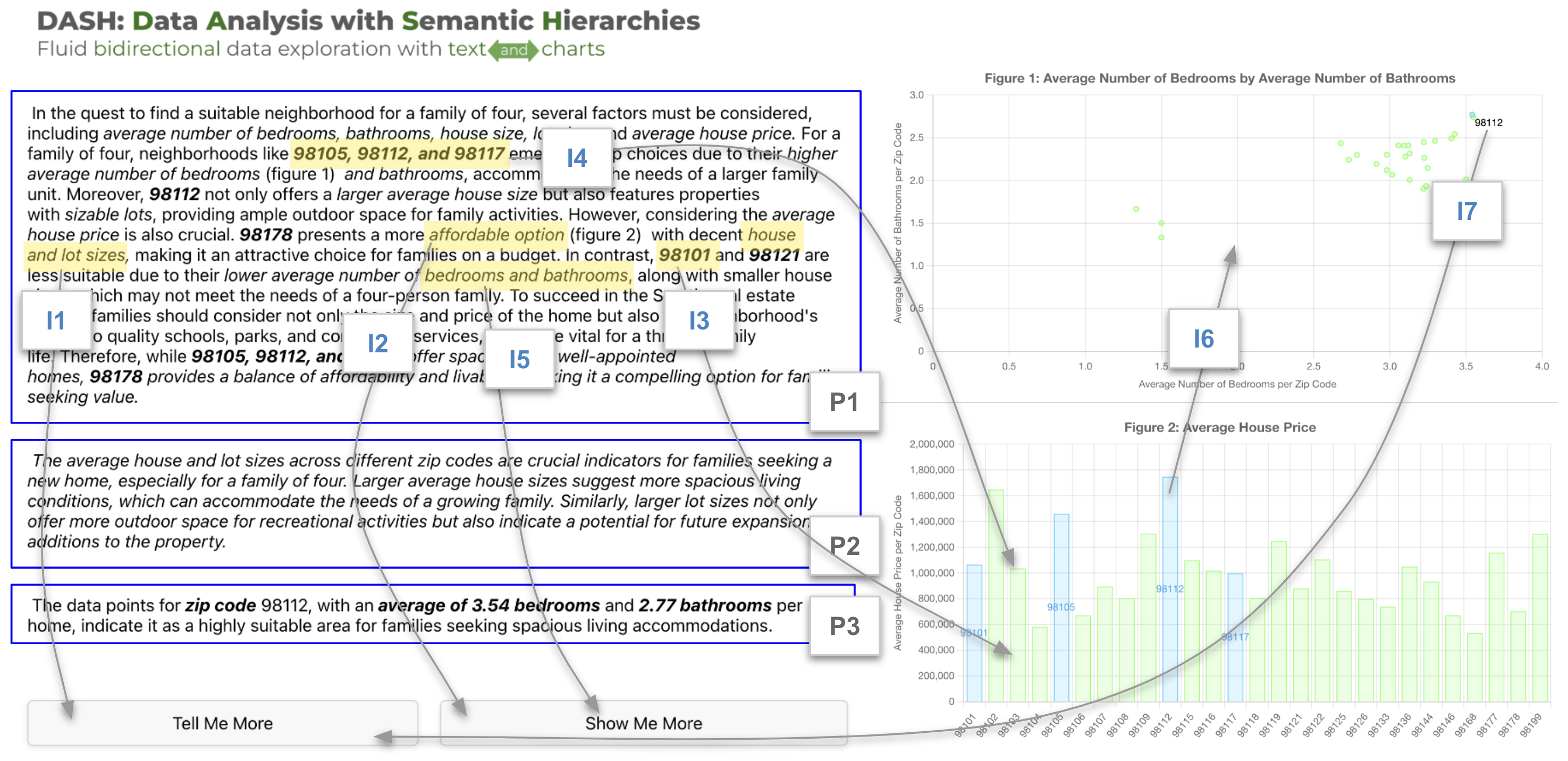}
\caption{The \dash~interface and an associated interactivity example. Generated text paragraphs are labeled as P\textit{x}, and interactive mouse-drag gestures are labeled as I\textit{x}. The \dash~interface comprises two primary sections: the text area on the left and the chart area on the right.  The \textit{Tell Me More} and \textit{Show Me More} buttons at the bottom left trigger additional text and chart rendering, respectively. Text-to-text (I1), text-to-chart (I2-I5), chart-to-chart (I6), and chart-to-text (I7) interactions are illustrated. 
}
\label{fig:dash-interactivity}
\end{figure*}

Figure~\ref{fig:dash-interactivity} shows an example of \dash~interactivity with generated text paragraphs (P\textit{x}) and interactions (I\textit{x}). \dash~presents an initial paragraph (P1) discussing a Seattle real estate dataset through the lens of a real estate agent looking for a good neighborhood for a family of four. The user drags the words ``house and lot sizes'' to the \textit{Tell Me More} button (I1).  \dash's LLM then generates a new high-level explanation of \textit{why} those attributes are important for their analytical goal (P2). The user then drags the phrase ``affordable option'' (whose metadata contains the `avg\_price' field) to the \textit{Show Me More} button (I2). \dash~generates the chart ``Average House Price'' and a chart reference is inserted into the text.  The user then drags the words ``98101'' and ``98105, 98112, and 98117'' into the ``Average House Price'' chart (I3 and I4), highlighting those zip codes.  The text ``bedrooms and bathrooms'' is then dragged to \textit{Show Me More} (I5), triggering \dash~to create a new scatter plot ``Average Number of Bedrooms by Average Number of Bathrooms''. Chart references are re-numbered by order of appearance in the text. Looking at the house price bar chart, the user wonders why zip code 98112 is so expensive, so they drag the ``98112'' bar from the bar chart to the scatter plot (I6), highlighting ``98112'' in the upper right of the scatter plot.  Looking for a higher-level analysis of what this position means, the user drags the ``98112'' mark to \textit{Tell Me More} (I7), whereupon the \dash~LLM generates text (P3) explaining that that zip code has high-end but expensive homes. The end result of this interaction is a custom bimodal dashboard tailored for a specific analytical end goal, where high-level analysis and low-level data presentation work together.

\section{Preliminary Evaluation}
To explore the utility of \dash~in supporting interactive data exploration, we conducted a preliminary user study involving 12 participants ($S1$ - $S12$) with diverse levels of experience in data analysis (six novices, two intermediate, and four experts). The primary goals were to: 1) gather feedback on the integration of semantic levels in textual and visual data exploration and 2) assess \dash's current design and identify areas for improvement.

Each session was conducted remotely, lasting about 30 minutes, with participants recruited via professional networks and social media platforms dedicated to data analytics communities. To mitigate any bias, participants were given only a brief introduction to the tool's capabilities at the beginning of the session. They were then asked to engage with \dash~ using the Seattle real estate data~\cite{SeattleRealestateData}, focusing on using the tool to generate data-driven narratives through the manipulation of text and visual elements. Participants performed the task of identifying a suitable neighborhood for a family of four by dragging and dropping data references across different semantic levels. This manipulation allowed participants to observe how actions changed the narrative and visualization context, providing insights into how well \dash~facilitates exploration toward the task goal. Feedback was collected through active observation and a semi-structured interview after the interactive session. 

\subsection{Results and Discussion}
Participants found \dash~to be a valuable tool for interactive data exploration in the context of the study task provided.

\pheading{Flexibility of interface for data exploration.} Participants generally found the interface of \dash~to be intuitive. They noted the ease with which they could manipulate data by dragging and dropping references across different semantic levels. $S1$ remarked, ``\textit{This feels very fluid, almost like moving pieces on a chessboard, but each piece reveals more information about the neighborhood."} The flexibility to dynamically alter the data narrative and visualization by adjusting semantic levels was appreciated by all the participants.

\pheading{Use of semantic layers to adapt the narrative.} \dash~was noted for its effective use of semantic layers to facilitate deeper insights into the data. Participants were impressed with the system's responsiveness and its ability to adapt the narrative and visualization based on their interactions. Challenges were noted when semantic layers were not immediately clear or when transitions between layers seemed abrupt, but overall, the system managed well. $S5$ noted, ``\textit{Seeing how my actions change the narrative and visual output in real time helps me understand the data at a much deeper level}."

\pheading{Relevance of data-driven text.} The relevance of the data-driven narratives to the task of finding a suitable neighborhood was consistently highlighted. Participants valued the tool’s capability to sift through data and present nuanced insights, such as identifying areas with good schools and low crime rates. For example, $S3$ excitedly shared, ``\textit{I dragged the average income level to the chart, and it immediately showed me areas that fit my family's lifestyle expectations}." $S11$ appreciated how \dash~could tailor the data presentation to reflect the task, stating, ``\textit{It’s impressive how it pulls together various data points to paint a clear picture of each neighborhood}."

\noindent Our evaluation also highlighted several areas for improvement: 

\pheading{Distinction across higher-level semantic levels.} Participants pointed out some limitations in the granularity expressed through the semantic layers. The interface did not always capture subtle distinctions within the data, which sometimes led to a generalized view rather than detailed insights. As $S7$ noted, ``\textit{When I tried to delve into more contextual data about local amenities, it felt like the system didn't quite capture every detail I was interested in}."

\pheading{Need for fine-tuning of generated content.} There was also feedback on the need for robust mechanisms to correct and refine data interactions. Participants expressed a desire for features in \dash~that allowed them to modify their interactions or fine-tune the semantic layers post-initial manipulation to better refine their exploration. $S6$ suggested, ``\textit{It would be helpful if I could tweak or even backtrack on the semantic layers after adjusting them and then feed it back to the system to fine-tune the information displayed}." Additionally, while the use of LLMs simplifies the implementation of \dash's bimodal flow, they are not without challenges. For example, during this project the LLM would sometimes mislabel text or inaccurately reference data.
Taken together with the performance and memory constraints of LLMs, where metadata labeling can take between 20-40 seconds, there are clearly areas for LLM improvement before they meet production tool standards. 

\section{Future Directions for Bimodal Data Analysis}
This preliminary work aims to operationalize the notion of semantic levels and demonstrate the feasibility of a fluid text/chart bidirectional analytical workflow. We identify several interesting directions for further enhancing the integration of text and charts in data exploration. One area for research is exploring techniques to enhance the customization of the semantic layers, providing users more flexibility to define and adjust these layers according to their domain-specific analysis needs. Future tools could incorporate machine learning models that not only interpret data but also predict trends and suggest new areas of exploration based on the user's interaction history and data patterns. This proactive suggestion system could guide users toward hidden insights and enhance their understanding of the data. Finally, integrating such tools with real-time data is important for wider adoption. Dynamically updating visualizations and text based on live data changes could be beneficial in environments where data is continuously evolving, such as in financial markets or social media analytics.


\bibliographystyle{abbrv-doi}

\bibliography{main}
\end{document}